\documentclass[aps,pre,twocolumn,groupedaddress,showpacs]{revtex4}
\usepackage{graphicx}

\begin{document}

\title{The Olami-Feder-Christensen earthquake model in one dimension}

\author{Felix Wissel and Barbara Drossel}
\affiliation{Institut f\"ur Festk\"orperphysik,  TU Darmstadt,
Hochschulstra\ss e 6, 64289 Darmstadt, Germany }

\begin{abstract}
We study the earthquake model by Olami, Feder and Christensen in one
dimension. While the size distribution of earthquakes resembles a power
law for small system sizes, it splits for larger system sizes into two
parts, one comprising small avalanches and showing a size independent
cutoff, and the other comprising avalanches of the order of the system
size. We identify four different types of attractors of the dynamics
of the system which already exist for very small systems. For larger
system sizes, these attractors contain large synchronized regions.
\end{abstract}


\maketitle

\section{Introduction}
\label{intro}

 The Olami--Feder--Christensen earthquake model \cite{ofc92} is
probably the most studied nonconservative and supposedly
self-organized critical model. Nevertheless, the origin of its
power-law like avalanche-size distribution is still not clear.  Apart
from these power laws, the model shows a variety of other interesting
and unusual features such as a marginal synchronization of neighbouring
sites driven by the open boundary conditions \cite{mid95}, and the
violation of finite-size scaling \cite{gras94,lis01} together with a
qualitative difference between system-wide earthquakes and smaller
earthquakes \cite{lis01a}. Also, small changes in the model rules
(like replacing open boundary conditions with periodic boundary
conditions \cite{per96}, or introducing frozen noise \cite{mou96}),
destroy the power laws. Recently, it was found that the results of
computer simulations are strongly affected by the computing precision
\cite{dro02}, and that the model exhibits sequences of foreshocks and
aftershocks \cite{her02}.

In order to better understand the model, we study here its
one-dimensional version. The model is highly nontrivial even in one
dimension, and some of its properties resemble those in two
dimensions. Just as in two dimensions, we find large synchronized
regions and a fundamental difference between the avalanches triggered
at the boundaries and those triggered deep inside the system. We
identify different types of attractors of the dynamics of the system
and explain the features of the model in terms of the properties of
these attractors. Our main finding is that the system in the
stationary state can be separated into a boundary region, where all
larger avalanches are triggered, and one (or two) synchronized inner
regions, the size of which can be varied without changing the
behaviour of the boundary region.

The outline of this article is as follows. In the next section, we
introduce the model rules.  In section \ref{small_systems}, we
focus on a system of up to 4 lattice sites and find its attractors. In
section \ref{formal}, we view the model from a dynamical systems'
perspective and present a general analytical approach that allows to
classify the attractors into four different types. In section
\ref{large}, we study larger systems. First, we investigate the
approach to the stationary state as function of the system size and
the model parameter. Then, we discuss the properties of the stationary
state. Finally, we summarize and discuss our main findings in section
\ref{conclusion}.

\section{The model}
\label{model}

The Olami-Feder Christensen model is a discretized and simplified
version of the Burridge-Knopoff model of earthquakes \cite{bur67}. In
an one-dimensional system consisting of $L$ sites, it is defined by the
following rules: At each site $i=1,\dots,L$, a continuous variable
$z_i$ is defined that represents a local force. The force at all sites
increases uniformly at a constant rate, which we set equal to 1. When
the force $z_i$ exceeds the threshold value $z_c$, which can be chosen
to be $z_c=1$ without loss of generality, the force at this site is
reset to zero, while the two nearest neighbours (or the only neighbour,
if the toppling site is a boundary site) receive a force increment of
$\alpha z_i$. The parameter $\alpha$ is the only parameter of the
model, and it has a value in the interval $(0,0.5)$. If a neighbour is
lifted above the threshold, the force on its neighbours is immediately
increased according to the same rule, etc., until the ``avalanche'' 
(the earthquake) is finished. 
The ``size'' of an avalanche, $s$, is defined to be the
number of toppling events during this avalanche. 
Such an earthquake is instantaneous on the time scale of driving. 
After the earthquake, the force is again increased with unit rate, 
until the next site reaches the threshold, 
triggering the next earthquake, and so on.

This model is deterministic, once the initial conditions are
given. Usually, the initial conditions are chosen randomly from a
uniform probability distribution for each site. Since the model is
deterministic and dissipative, it has attractors of the dynamics.

From a dynamical systems' perspective, the model can be viewed as a
$L$--dimensional map, which maps the state of the model after one
avalanche (which may have size 1) on the state after the next
avalanche. Due to the toppling, the map is non continuous.

If $\alpha > \alpha_c=(\sqrt{3}-1)/2\simeq 0.366$, a site that topples
can in principle receive from its neighbours packages of a total size
larger than 1, causing the first site to topple again. Throughout this
paper, we assume that each site topples only once during an avalanche,
and we limit therefore our numerical studies and analytical arguments
to the case $\alpha<\alpha_c$, except where indicated otherwise.

Let us briefly summarize a few known results that are relevant for our
study of the one-dimensional system. 
Firstly, it is found that for
periodic boundary conditions the dynamics approach an attractor where
every site gives to its neighbours only force packages of size
$\alpha$. 
This means that no force value exceeds the threshold value
$z_c$ on the attractor. After a time $1-2\alpha$, each site has
toppled once and has received two force packages of size $\alpha$ from
its neighbours. This means that after time $1-2\alpha$ the force on
each site is again the same. Slightly increasing or decreasing force
values gives again a periodic orbit with period $1-2\alpha$, as long
as this change does not cause a toppling site to lift its neighbour
above the threshold.

Secondly, the open boundary conditions are responsible for the
occurrence of large avalanches and large synchronized regions, where
neighbouring sites differ in force values only by a small amount. A
nice explanation for this has been suggested by Middleton and Tang ten
years ago \cite{mid95}. They considered a system of two sites where
one site is driven at a slower rate than the other. This mimicks the
fact that sites close to the boundary receive on an average less force
packages than those deep inside the system. The two sites settle on an
attractor where the slower site always topples first and lifts the
faster site above the threshold. The faster site therefore looses more
force when it topples, and the slower site receives a larger force
package. This compensates the different driving speed, and the two
sites remain synchronized and always topple together.

Thirdly, the largest possible force package that a site can pass to
its neighbour is $\alpha/(1-\alpha)$. This package size is reached if
an avalanche passes through a region where all sites are at the
threshold.

\section{Very small systems}
\label{small_systems}

\subsection{$L=2$}
\label{l2}

For $L=2$, any state where the force 
difference between the two
neighbours is larger than $\alpha$, is part of a cycle of period
$1-\alpha$. After a time $1-\alpha$, each site has toppled once and has 
received a force package of size $\alpha$ due to the toppling of the
neighbour.  Furthermore, it has received a force increment $1-\alpha$
due to the uniform driving. The net change in force is therefore zero.

\subsection{$L=3$}
\label{l3}

For $L=3$, we consider the state of the system whenever the middle
site has toppled, i.e., when $z_2=0$.  $z_1$ and $z_3$ then take
values in {$[\alpha, 1+\alpha +\alpha^2]$} only.  The lower value
$\alpha$ is realized, if a site just toppled itself and
lifted $z_2$ exactly to the threshold.  The upper limit occurs only
if all sites were at the threshold before. 

\begin{figure}[htb]
\begin{center}
\mbox{}
\includegraphics[width=\columnwidth]{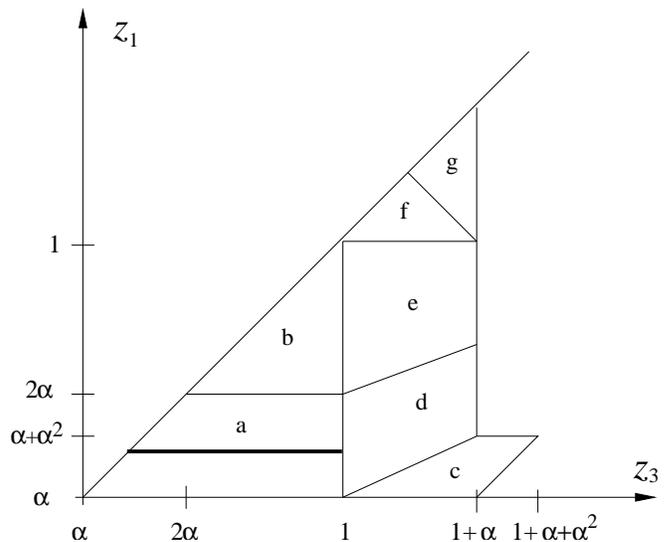}
\end{center}
\caption{\label{figure1} Different regions in the $z_1$--$z_3$ plane
(with $z_2=0$) that have different toppling sequences. Only the part
with $z_1 \le z_3$ is shown. Thin lines separate regions, the thick
line is the fixed point line. }
\end{figure}

Without loss of generality, we assume $z_1 \le z_3$.  Different
regions in the $z_1$--$z_2$ plane can be characterized by their
sequence of topplings $t_i$ and of growth, until $z_2=0$ again.  An
example of this would be $(t_3,g,t_1,t_2)$, for which first the right
site topples, followed by a uniform growth, and then the toppling of
the left site lifts $z_2$ over the threshold.  There is a total of 14
different such regions, seven of which are marked by letters $a$ to
$g$ in figure \ref{figure1}.

By investigating the transitions between these regions, one finds two
attractive fixed point lines. One fixed point line is at
$z_1=z^*=\frac{\alpha(1+2\alpha)}{1+\alpha}$ with $z_3 \in
(z^*,1]$. The other fixed point line is obtained by interchanging
$z_1$ and $z_3$.
The corresponding attractor is a cycle of two avalanches, written as
$(g,t_j,g,t_i,t_2)$ in the above notation, with $i$ being the site with
$z_i=z_i^*$, and $j$ being the other neighbour of site $2$. 

A special case is the symmetric case $z_1=z_3=z^*$.
Site 1 and 3 topple simultanously, 
thereby lifting site 2 above the threshold,
and receiving a package $z^*$ in turn.

To summarize, a system with $L=3$ approaches a periodic attractor with
2 different avalanches.

\subsection{$L=4$}
\label{l4}
\begin{figure}[htb]
\begin{center}
\mbox{}
\hspace*{-1cm}\includegraphics[width=\columnwidth]{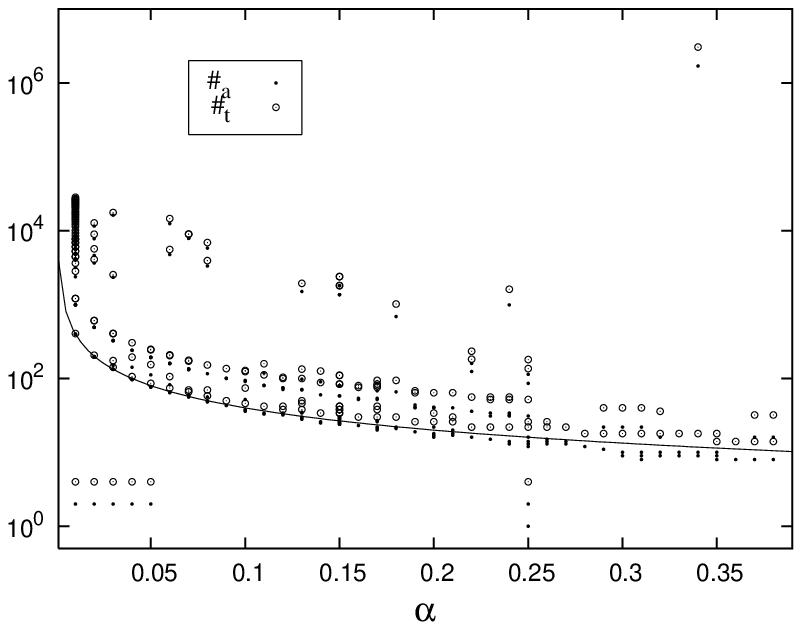} 
\end{center}
\vspace*{-0.8cm}
\begin{center}
\hspace*{-1cm}\includegraphics[width=\columnwidth]{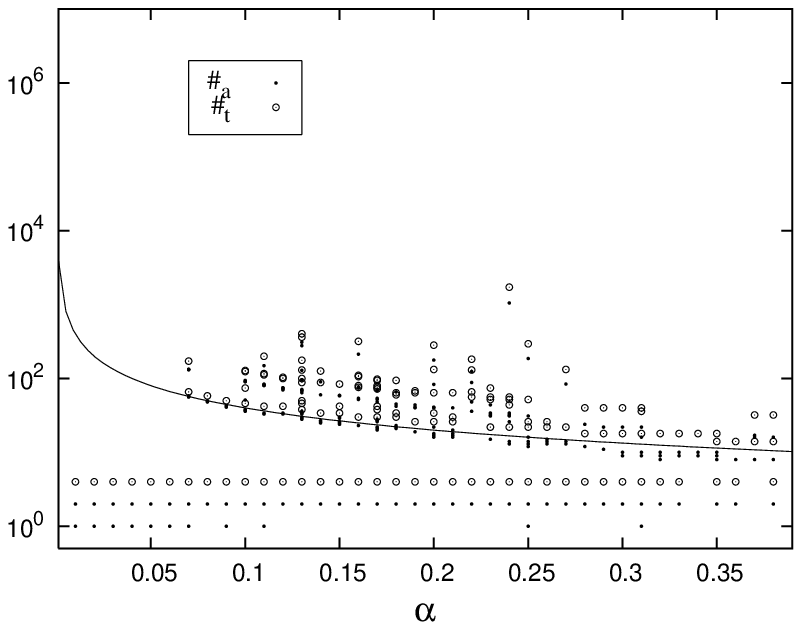}
\end{center}
\caption{\label{figure2} Period of attractors as function of
$\alpha$, measured in the number of avalanches per period 
$\#_a$ (dots) and in the number of topplings per period $\#_t$ (circles). The
solid line corresponds to $f(x)=4/x$. Top: precision $2^{-20}$;
bottom: precision $2^{-10}$. The points above $10^6$ for
$\alpha =0.34$ are valid data points.
}
\end{figure}

For the system size $L = 4$, all attractors of the dynamics are
periodic. We find a variety of different attractors for a given value
of $\alpha$. Figure \ref{figure2} shows the period of the attractors
found doing a scan of the same $128^4$ different initial conditions
for each value of $\alpha$ and for 2 different precisions.  The period
is measured once in terms of the number of topplings, $\#_t$, and once
in terms of the number of avalanches, $\#_a$.

One can discern the following features:
\begin{enumerate}
\item Degeneracy: Attractors with different numbers of avalanches per
period have the same total number of topplings per period. We found
that different attractors can have a different
toppling sequence, while the force packages that each sites receives
from its neighbours are identical. One explanation for this is that
there are mechanisms in the model that create a degeneracy of
different sites, for instance when two sites have been the end points
of the same avalanche (after which they both had zero force) and
remain synchronized until they reach the threshold again. The toppling
sequence then depends on the exact implementation of the algorithm and
on rounding errors due to finite precision; one or the other version
of the attractor can be obtained depending on the initial
conditions. The total number of topplings and the packages sizes can
be identical on the two versions of the attractor if there is a site
between the two synchronized sites that topples only after having
received a package from both sides.
\item Persistence: Attractors exist over a certain interval of
$\alpha$ values. If $\alpha$ is changed slowly (such that the system
can follow adiabatically), the attractor remains the same as long as
the avalanches remain the same. Eventually, a point will be reached
where an avalanche decays into two avalanches (because the size of a
package is no more large enough to lift the neighbour over the
threshold), or where two avalanches merge to form one avalanche
(because the distance between two neighbours has become smaller than
the package size). As we will see in the next section, the stability
of the attractor can change at this point (but typically not before). If the
state with the new toppling pattern is stable, we have a new attractor
with the same period (if measured in number of topplings). Otherwise,
if the new state is unstable, the system moves to a different
attractor, and we obtain a step in Figure \ref{figure2}.
\item Divergence of the periods at small $\alpha$: At small $\alpha$,
the period of the attractors is increasing. This can be explained by
considering a boundary site and its neighbour. During each time
interval $1-2\alpha$, the boundary site receives a package of the
order of $\alpha$ from its neighbour, while the neighbour receives two
packages of the order of $\alpha$. It therefore takes of the order of
$1/\alpha$ time intervals until the two sites have again roughly the same
height.  For this reason, the period diverges as $1/\alpha$.
\item Smallest possible period: For all values of $\alpha$, the
shortest attractor has 4 topplings.  In configuration space, this
corresponds to the state $(\tilde{z}, \tilde{z},0,\alpha+\alpha^2)$,
with an arbitrary force $\tilde z \in [\alpha+\alpha^2, 1]$, and with
the toppling sequence $g,t_1,t_2,(g),t_4,t_3$. Whether this attractor
is realized, depends on the implementation of the algorithm.  For
smaller $\alpha$ and for smaller precision, it occurs more often
and eventually has the weight one.
\item Vastly different periods: For a given $\alpha$, there exist
attractors with widely different periods. The most prominent periods
lie in two bands, which are clearly visible in the figure. For
certain values of $\alpha$, very large periods occur with a
considerable weight. Attractors with these large periods typically
have toppling sequences that are most of the time periodic with a much
shorter period than that of the attractor, but the force values do not
have this short period.
\item Sensitivity to the computing precision: Attractors with larger
periods occur (for the same value of $\alpha$) less often when the
computing precision is smaller. 
The reason for this is that on longer attractors,
there are more states at all 
and states that are close to each other are more likely to occur.
In a simulation with smaller numerical precision, such
states can become identical, and the period of the attractor becomes
shorter.
\end{enumerate}

\section{Analytical approach and classification of attractors}
\label{formal}

We describe the state of the system as a difference vector $\vec x =
(x_1,\dots,x_{L-1})$ with $x_i = z_{i+1}-z_i$. The uniform increase in
force does not change this vector, but toppling sites do change
it. The force value of the first site toppling in an avalanche is
decreased by one, 
while its two neighbours receive a package $\alpha$.
This process can be decribed by adding a vector
$$(0,\dots,0,-\alpha,1+\alpha,-(1+\alpha),\alpha,0,\dots,0)$$ to $\vec
x$ (with the four nonzero elements at the appropriate place).  In
contrast, a subsequent toppling event can be described by applying to
$\vec x^t$ a matrix that is identical to the unity matrix everywhere
except for a $4\times 4$ block on the diagonal.  There are two
different matrices, corresponding to avalanches propagating to the
right and to the left.  We assume that site $i$ just toppled.  Then,
the next toppling event is represented by the matrix
\begin{equation}
\mathbf{M}_l^{i-1} = 
\left(
\begin{array}{cccccccc}
1&0&\dots&&&&&0\\
0&\ddots&&&&&&\vdots\\
\vdots&&1&0&-\alpha&0&&\\
&&0&1&1+\alpha&0&&\\
&&0&0&-\alpha&0&&\\
&&0&0&\alpha&1&&\\
&&&&&&\ddots&0\\
0&\dots&&&&&0&1
\end{array}
\right) \label{mrl}
\end{equation}
if the avalanche is moving to the left. The nontrivial column (number $i-1$)
describes the toppling of site $(i-1)$, which was lifted above 
the threshold by the prior toppling of site $i$. 

If the avalanche is moving to the right,
the corresponding matrix reads 
\begin{equation}
\mathbf{M}_r^{i+1} = 
\left(
\begin{array}{cccccccc}
1&0&\dots&&&&&0\\
0&\ddots&&&&&&\vdots\\
\vdots&&1&\alpha&0&0&&\\
&&0&-\alpha&0&0&&\\
&&0&1+\alpha&1&0&&\\
&&0&-\alpha&0&1&&\\
&&&&&&\ddots&0\\
0&\dots&&&&&0&1
\end{array}
\right) \label{mrr}
\end{equation}
with the nontrivial column being number $i$. If site $i-1$ and site
$i+1$ both topple simultaneously, the matrix contains both nontrivial
columns.  If the toppling site is a boundary site or the site next to
it, surplus columns and rows of the nontrivial block of
$\mathbf{M}_l^i$ or $\mathbf{M}_r^i$ are to be removed.

We now focus our interest on the difference $\delta \vec x$ between
two systems and assume that the two systems have the same toppling
sequence. This will be the case as long as they are sufficiently close
to each other. Then they will both be updated by adding the same
vectors and multiplying the same matrices in the same order. Adding
the same vector to the state $\vec x$ of both systems has no effect on
the difference $\delta \vec x$. The difference between the states of
the two systems will therefore evolve solely by multiplying matrices
of the form $\mathbf{M}_l^i$ and $\mathbf{M}_r^i$ to it (-- except if
the first toppling site lifts both neighbours above the threshold; in
this case the first matrix associated with this avalanche contains two
nontrivial columns, while the other matrices have the usual form,
since the two branches of the avalanche commute after the first toppling.)

Let one of the two systems be on a periodic orbit. Whether the other
system will approach the orbit, depends on the largest
eigenvalue of the product
\begin{equation}\mathbf{S}=\mathbf{M}_{\nu(P)}^{i(P)}\mathbf{M}_{\nu(P-1)}^{i(P-1)}\dots\mathbf{M}_{\nu(2)}^{i(2)}\mathbf{M}_{\nu(1)}^{i(1)}
\label{product}
\end{equation}
with $P$ being the total number of matrices occurring during one
period, $i(p)$ being the nontrivial column index for matrix number
$p$, and $\nu(p)$ being $l$ or $r$ depending on whether an avalanche
is moving to the left or to the right.

If the largest eigenvalue of $\mathbf{S}$ is larger than one, the
orbit is unstable and cannot be an attractor of the dynamics. In the
following, we describe four types of attractors that we have found in
the model. All these attractors occur already in a small system of
$L=4$, but are also seen in large systems.

\subsection{Marginally stable attractors}
\label{marginally}
Marginally stable attractors occur if the largest eigenvalue of
$\mathbf{S}$ is identical to 1. We observed regularly attractors where
a column $i$ of $\mathbf{S}$ is identical to the unit vector $\vec
e_i$. This means that $\vec e_i$ is an eigenvector of $\mathbf{S}$,
and adding a small multiple $\epsilon \vec e_i$ to the periodic orbit
gives again a periodic orbit with the same toppling sequence. In terms
of forces $z_j$ this means that increasing or decreasing all force
values $z_j$ with $j \le i$ by a small amount results again in a
periodic orbit. The product (\ref{product}) contains no matrix
$\mathbf{M}_{l}^{i}$ or $\mathbf{M}_{r}^{i+1}$. Sites $i+1$ and $i$
never cause each other to topple. There is no avalanche that includes
simultaneously site $i+1$ and site $i$. We say that there is a
``barrier'' between sites $i+1$ and $i$. We found that the total size
of the force packages that site $i+1$ gives to site $i$ during one
period is identical to the total size of the force packages that site
$i$ gives to site $i+1$. For $L=4$, the barrier is always in the
middle of the system. For larger sizes, it need not be in the middle,
but it is often found at the center of a system, since the
synchronization proceeds at constant speed from the boundaries (see below).

Sites $i$ and $i+1$ to the right and left of the barrier must topple
equally often during one period. If this was not the case and if site
$i$ toppled more often, there would be an instance where site $i$
topples twice without site $i+1$ toppling in between. After site $i$
has toppled for the first time, its force is zero, and that of site
$i+1$ is at least as large as $\alpha$. In order for site $i$ to reach
the threshold before site $i+1$, it must receive a package from its
left neighbour that is larger than $\alpha$, while site $i+1$ receives
no package. The largest possible package size is $\alpha/(1-\alpha)$,
and therefore site $i+1$ has at least the force
$1-\alpha^2/(1-\alpha)> 1-\alpha$ at the moment where site $i$ reaches
the threshold for the second time. This means that site $i+1$ is
lifted above the threshold by the toppling of site $i$, in
contradiction to our assumption that there is a barrier between the
two sites. Therefore, the two sites must topple equally often.

There exists no attractor with 2 or more barriers. We show this in two
steps. First, let us assume that all force packages passed over the
barriers are of size $\alpha$. Then the region between the two
barriers is like a system with periodic boundary conditions, and no
package passed on within this region is larger than $\alpha$.  The two
sites immediately outside the barriers must not be lifted above the
threshold by their neighbours. Otherwise, they would pass packages
larger than $\alpha$ over the barrier. Furthermore, the two sites
immediately outside the barriers must not topple more often than the
sites between the barriers. Therefore they cannot receive packages
larger than $\alpha$ from their outward neighbours. This means that the
sites immediately outside the barriers in fact also belong to the
domain of sites that are never lifted above the threshold and that
always receive packages of size $\alpha$. By repeating this argument,
we find that no site in the system can be lifted above the
threshold. However, this situation cannot be realized with open
boundary conditions. It occurs for periodic boundary conditions.

Now, since we have ruled out the possibility that the region between
the two barriers receives only force packages of size $\alpha$ from
outside, let us assume next that they receive on an average packages
larger than $\alpha$. We simulated the region between the two barriers
by inserting packages of a size larger than $\alpha$ at its boundaries
immediately after the boundary sites have toppled. This leads to
attractors where avalanches are triggered at the center of the system
and are running outwards. The attractors and therefore the number of
topplings per unit time of the boundary site are determined by the
size of the region and the size of the packages received from
outside. On the other hand, this number of topplings must be identical
to the number of topplings of the site on the other side of the
barrier in the original system. However, there is no free continuous
parameter left to match this condition, and it can therefore usually
not be satisfied. This problem does not arise in the case of a single
barrier, because the state of the system can be symmetric about the
barrier, thus satisfying the matching conditions. 

Finally, let us consider a periodic orbit at the boundary of the basin
of attraction of the marginally stable attractors. In order to obtain
this orbit, we increase or decrease all force values $z_j$ with $j \le
i$ by an amount such that there is a moment in time where site $i$ (or
$i+1$) is lifted by site $i+1$ (or $i$) exactly to the threshold. The
metastable orbit has now become degenerate with an orbit where site
$i$ (or $i+1$) is lifted by site $i+1$ (or $i$) infinitesimally above
the threshold. This orbit has no barrier, and the matrix $\mathbf{S}$
corresponding to this periodic orbit is different from the one
corresponding to the metastable orbit. Its largest eigenvalue will
therefore be different from 1. We have seen realizations of the
interesting case that the largest eigenvalue becomes smaller than
1. This means that the periodic orbit that is at the boundary of the
basin of attraction of the marginally stable attractors can itself be
an attractor that is reached from a nonvanishing set of initial
conditions. 

\subsection{Strongly stable attractors}

In general, 
the matrix $\mathbf{S}$ is obtained by starting with the unit matrix
and multiplying the appropropriate matrices $\mathbf{M}_{l}^{i}$ or
$\mathbf{M}_{r}^{i}$ one after the other. 

The effect of $\mathbf{M}_{l}^{i}$ (or $\mathbf{M}_{r}^{i}$) on a
matrix is that row number $i$ (or row number $i-1$),
multiplied by a certain factor $\pm\alpha$ or $(1+\alpha)$,
is added to 3 neighbouring rows
and is itself multiplied by $-\alpha$.

Let us perform an expansion in powers of $\alpha$ and focus
on the elements of order $\alpha^0$. 
In order $\alpha^0$, the matrices
$\mathbf{M}_{l}^{i}$ and $\mathbf{M}_{r}^{i}$ simply add one row to a
neighbouring one and replace the original row with all zeros. 
If a boundary site is caused to topple by its neighbour, all 1's that have
been in the boundary row are flushed out of the system. 
Starting with a unit matrix, all 1's that remain in the system are in the 
same row after multiplying a sufficient number of 
$\mathbf{M}_{\nu}^{i}$ matrices. 

We have never seen an attractor where this does not happen. Any given
site of the system is reached by an avalanche that starts near the
boundary, and therefore there cannot be 1's left in different
rows. The marginally stable attractors have all 1's in a row $i$,
where they stay forever.

However, there is also the possibility that all the 1's are flushed
out of the system by  avalanches that extend from inside the system
to the boundary. In this case, the largest eigenvalue of $\mathbf{S}$
is of the order $\alpha$, and the attractor is quickly approached. We
have seen many examples of such strongly stable attractors.

\subsection{Weakly stable attractors}

If not all 1's are flushed out of the system, and if there is no
barrier in the system, the largest eigenvalue of $\mathbf{S}$
belonging to an attractor is $1-\mathcal{O}(\alpha^n)$ with some power
$n$ of $\alpha$.  This corresponds to the situation where the row
containing the 1's remains in the system and is moved around by
avalanches coming from both directions.

If $\alpha$ is small or $n$ is large, attractors are
approached very slowly. For $\alpha=0.2$, we have seen attractors with
$n=2$ for $L=4$ and an estimated $n=11$ for $L=20$. 
However, we did not attempt a systematic survey of relaxation times 
towards the attractors as function of $\alpha$ and $L$.

\subsection{Complex attractors}

There exist attractors with strikingly long periods of the order of
many thousands for $L=4$ or much larger for larger $L$ (if period is
measured in total force increment per site). Typically, these
attractors contain long quasiperiodic sections where the sequence of
avalanches remains the same but the force values change slowly, just
as one would expect close to a weakly stable or weakly unstable
periodic orbit. This quasiperiodic sequence is eventually interrupted
by an intermittent phase containing other avalanches, until the
quasiperiodic phase is entered again. One can understand the origin of
such complex attractors in the following way. Imagine a weakly stable
attractor for a certain value of $\alpha$. Now change $\alpha$ slowly
and let the system follow adiabatically. The largest eigenvalue of
$\mathbf{S}$ on the resulting attractor will have the same
coefficients if expanded in powers of $\alpha$, as long as the
avalanches remain the same. Eventually, a value of $\alpha$ will be
reached where two avalanches merge or an avalanche splits, changing
the product of $\mathbf{M}$ matrices, which now can have an eigenvalue
larger than 1. Nevertheless, there may still be a region nearby in
state space where the old sequence of avalanches can be maintained for
a long time if the largest eigenvalue of $\mathbf{S}$ of the old
attractor was only slightly smaller than 1 (i.e. if the old attractor
was weakly stable). 

\section{Numerical investigation of large systems}
\label{large}

\subsection{Transient stage}
\begin{figure}[htb]
\begin{center}
\mbox{}
\includegraphics[width=0.48\columnwidth]{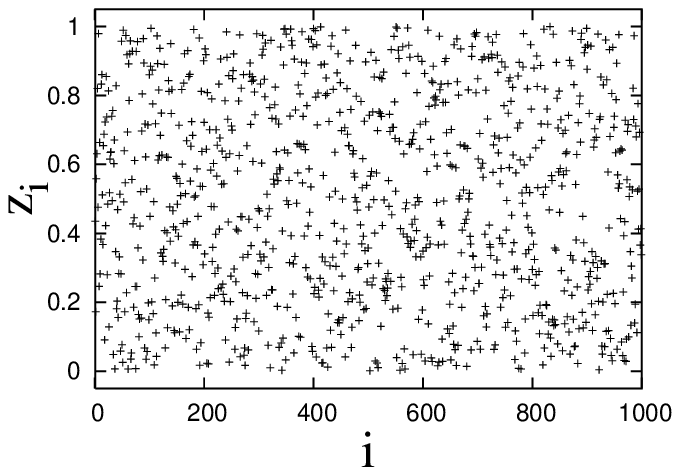}
\includegraphics[width=0.48\columnwidth]{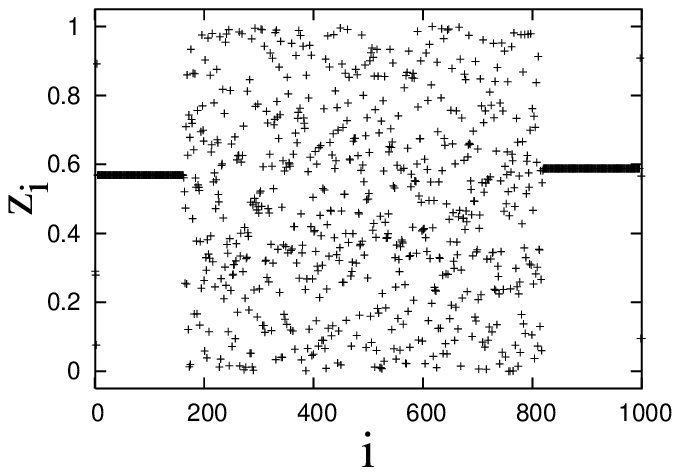}
\vspace*{0.2cm}

\includegraphics[width=0.48\columnwidth]{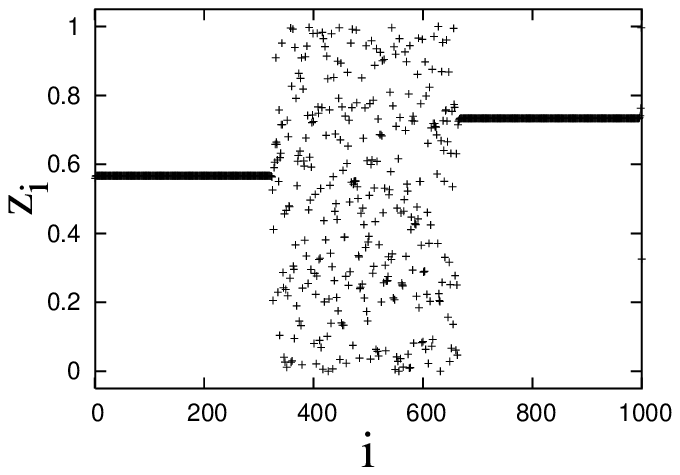}
\includegraphics[width=0.48\columnwidth]{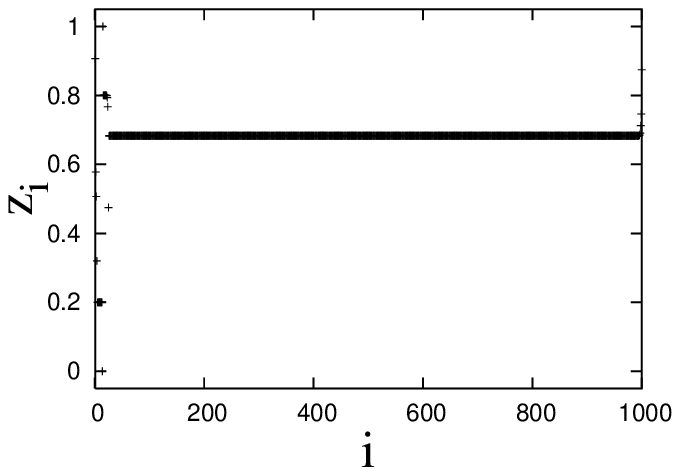}
\end{center}
\caption{\label{figure3} 
Random initial configuration and 
after $3\times 10^6$,
$5\times 10^6$ and $9\times 10^6$ avalanches in configuration space 
for $L=1000$ and $\alpha=0.2$ 
(from top left to bottom right).
}
\end{figure}

We now present simulation results for larger systems. We first study
the transition from a random initial state to a stationary
state. 
Figure \ref{figure3} shows the force values throughout a system of size
$L=1000$ for different times. 
One can see that starting from the boundary
more and more sites become synchronized, until in the stationary state
all sites apart from a few sites at the boundary have almost the same
force $z_i$. 

We have evaluated the transient time using two different measures of the
degree of synchronization:
(i) The standard deviation
$$\sigma^2(t)=\min_{0<\tau\leq t}\left(\bar{z}^2(\tau)-\overline{z^2}(\tau)\right)$$
as function of $\alpha$ and
$L$, where the bar denotes the average taken over all sites $i$.
(ii) Nearest-neighbour deviation 
$$\sigma_{NN}(t)=\min_{0<\tau\leq t}\left(
\overline{z_{NN}}(\tau)-\overline{z^2}(\tau)\right)$$ 
where again the bar denotes the average over all $i$ and 
$\overline{z_{NN}}=\frac{1}{N-1}\sum_{i=1}^{N-1} z_i\, z_{i+1}$
Time was measured as the number of topplings per site 
(i.e. the total number of topplings divided by the system size).

\begin{figure}[htb]
\begin{center}
\mbox{}
\includegraphics[width=0.48\columnwidth]{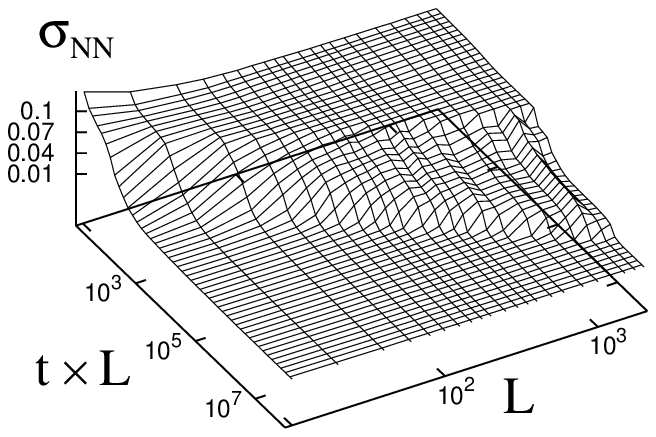}
\includegraphics[width=0.48\columnwidth]{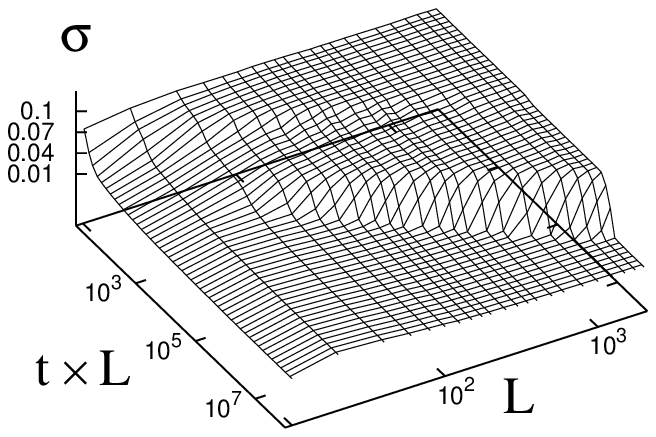}
\includegraphics[width=0.48\columnwidth]{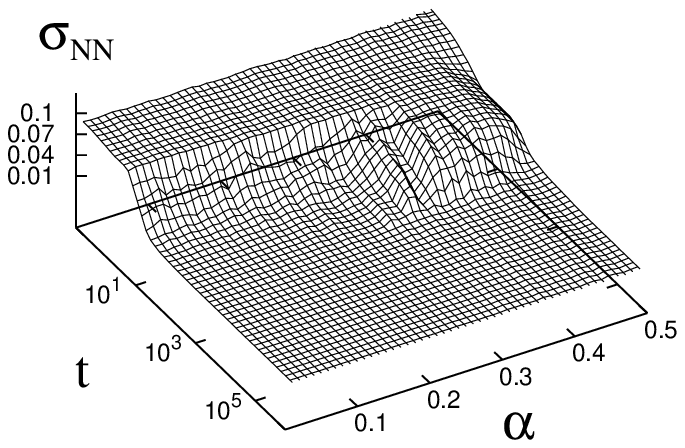}
\includegraphics[width=0.48\columnwidth]{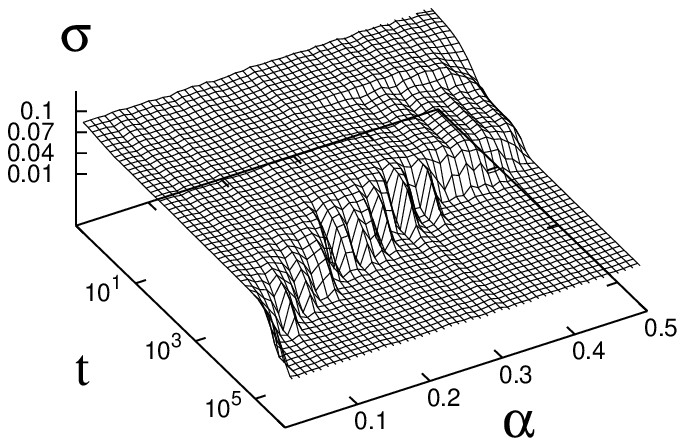}
\end{center}
\caption{\label{figure4}
Standard deviation $\sigma$ (right) and nearest neighbour deviation $\sigma_{NN}$
(left) for fixed value of $\alpha=0.35$ (top) and fixed value of $L=200$ 
(bottom) as function of time (measured in number of topplings per site) and
 $L$ and $\alpha$ respectively, note that time is plotted in log-scale 
and the double logarithmic scale in the upper two figures.}
\end{figure}

 Figure \ref{figure4} shows our results for these two
synchronization measures, averaged over 1000 random initial
configurations.  The two figures at the top show the behaviour of
$\sigma_{NN}$ (left) and $\sigma$ (right) as a function of time and of
the system size $L$.  The lower plateau indicates clearly the
stationary state. One finds that the transient time is proportional to
the system size.  This is due to the inward proceeding
synchronization, which takes place at a constant rate, if $L$ is
sufficiently large. For small $L$, the boundary layer takes a large
part of the system, and there is therefore little synchronization.
Apart from the transition to the stationary state with a small value
of $\sigma$ and $\sigma_{NN}$, one can also distinguish an earlier
transition, where the two measures leave the value $1/12$
corresponding to a random initial configuration. We interpret this
transition as the onset of the formation of synchronized blocks, after
the boundary layer has been set up. The characteristic shape of the
curves between these two transitions is strikingly different.  While
the nearest-neighbour deviation decreases rather fast once the
synchronization starts, the standard deviation remains on a second
plateau until its final decrease.  This comes from the two
synchronized blocks, which usually have a different force value, as
one can see in the second picture of figure \ref{figure3}.

\begin{figure}[htb]
\begin{center}
\mbox{}
\includegraphics[width=0.8 \columnwidth]{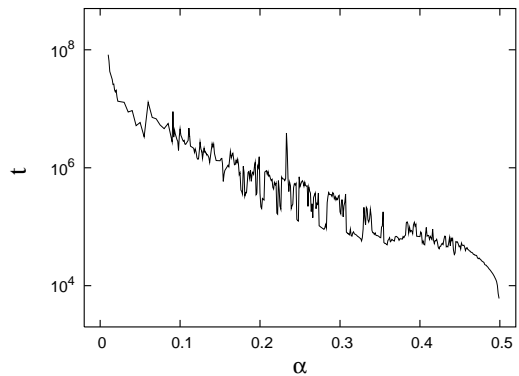}
\end{center}
\caption{\label{figure5} Time (measured in topplings per site) needed
to reach $\sigma=0.01$ for a system of $L=200$ sites as a function of
$\alpha$, averaged over 1000 different random initial systems.  The
peaks are real and are not statistical fluctuations.  }
\end{figure}

\begin{figure}[htb]
\begin{center}
\mbox{}
\includegraphics[width=0.8 \columnwidth]{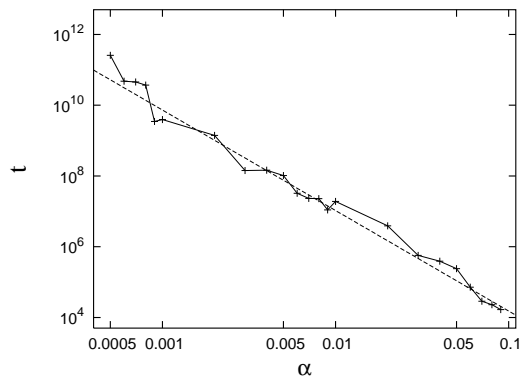}
\end{center}
\caption{\label{figure6} Time (measured in topplings per site) needed
to generate the first avalanche of size $\ge 21$ as function of
$\alpha$, averaged over 1000 different random initial states.  The
straight line corresponds to $f(\alpha)\propto\alpha^{-2.84}$.}
\end{figure}

A similar behaviour is found for the dependency on $\alpha$ as shown
in the lower two figures for a fixed system size $L=200$.  For values
of $\alpha$ below the critical value $\alpha_c$, the nearest-neighbour
deviation depends only weakly on $\alpha$. The sharp transition at the
end of the high plateau of $\sigma_{NN}$ is linear in $\alpha$. The
bulky part above $\alpha_c$ must be due to the fact that a site can
now topple twice during the same avalanche. While the onset of
synchronization is better visible in the data for $\sigma_{NN}$, which
decay very rapidly, the transition to the stationary state is much
better visible in the data for $\sigma$, which remain close to the
initial value for a longer time. A more detailed investigation of the
transition time to the stationary state reveals the following: (i)
Over a wide range of $\alpha$ values this transition time depends
exponentially on $\alpha$, as shown in Figure \ref{figure5}, where the
time is plotted that is needed to reach $\sigma=0.01$. (ii) For small
values of $\alpha$, the data show a power law with an exponent around
${-2.84}$ (Figure \ref{figure6}). Since the synchronization proceeds
very slowly, we did not measure the time to the stationary state, but
the time to show for the first time an avalanche larger than 20.
The following analytical arguments suggest that the transient time
should indeed diverge at least as fast as $\alpha^{-2}$ with
$\alpha$. Let us define a time unit as the time during which a force
$1-2\alpha$ is added to the system. A site at the center receives two
packages of size $\alpha$ from its neighbours per unit time and
topples on an average once per unit time. A boundary site receives
only one package and has on an average approximately $1-\alpha$
topplings per unit time. A site in the synchronized block topples on
an average $y$ times per unit time, with $y$ being intermediate
between these two limit cases, $1-\alpha < y < 1$. Initially, the
force difference between the synchronized block and the site that will
be synchronized next, is of the order of 1. In order to decrease this
force difference to a value of the order of $\alpha$, the difference
in the total number of topplings between the block and its neighbour
must be of the order of $1/\alpha$, which is achieved after a time of
the order of $1/\alpha(1-y)$. This increases with decreasing $\alpha$
at least as fast as $1/\alpha^2$.

\subsection{Stationary state}
\label{stationary}

We now turn to systems already in the stationary state and give an
overview of their statistical behaviour.

The most striking feature of the stationary state are the large
synchronized blocks.  Sites within the blocks topple the same number
of times, while sites closer to the boundaries topple less often. From
time to time, a large avalanche that begins outside the block runs
through the entire block. Between the large avalanches, the sites
within a block topple mostly one by one, lifting each other almost
exactly to the threshold. Let us consider a small region within such a
synchronized block and let us show that the sites in this region must
have approximately the same height, given the dynamics just
described. When the sites topple one by one, their height differences
are exactly the same as before, after each site has toppled once. When
an avalanche enters the region from  outside and extends
several sites beyond it, the change in height differences is
calculated by multiplying the state vector $\vec x$ with the
appropriate product $\mathbf{S}^{av}$ of $\mathbf{M}_{\nu}^{i}$ matrices. 
If the avalanche passes our region from the right to the left, the
elements of $\mathbf{S}^{av}$ in a row $i$ belonging to our region are
\begin{equation}
\mathbf{S}^{av}_{ij} = 
\left\{ 
\begin{array}{ll}
\alpha^{2+j-i} & \mbox{ for } \; i-1\le j \le
j_{ini}\\
0 & \mbox{ else }
\end{array}
\right.
\end{equation}
 with $j_{ini}+1$ being the site that triggered the avalanche.
If we denote with $x_i$ and  $x_i'$ the values of the force differences 
within our region before and after the avalanche, we have
$$x_i'=\alpha(x_{i+1}'+x_{i-1})< 2\alpha \mbox{ Max}\;
(x_{i+1}',x_{i-1}) $$ for an avalanche passing through the region from
the right to the left.  The asymptotic values of the $x_i$ after many
avalanches satisfy $x_i = \alpha(x_{i+1}+x_{i-1})$ within the
synchronized region. This condition can only be satisfied with all
$x_i$ being zero or with $x_i$ decreasing by a factor of the order
$\alpha$ from one value of $i$ to the next. Deep inside the
synchronized region, the $x_i$ become therefore very small.

\begin{figure}[htb]
\begin{center}
\mbox{}
\includegraphics[width=\columnwidth]{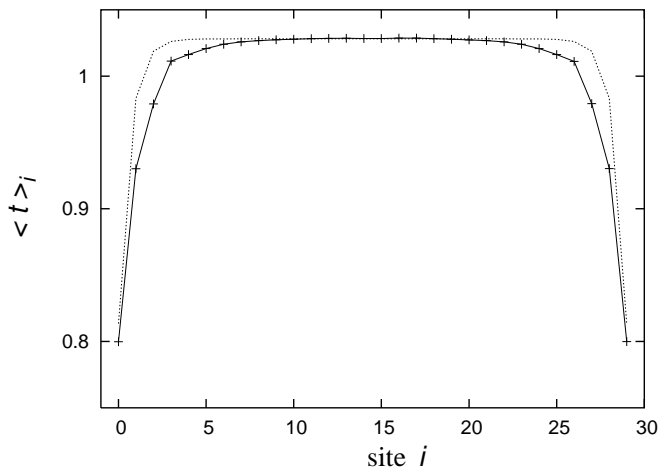}
\end{center}
\caption{\label{figure7}
Thick line: mean number of topplings $\langle t\rangle_i$ as a function of site index $i$, 
averaged over 1000 different synchronized systems and $10^4$ topplings
for $\alpha=0.2$. 
dotted line: analytical result for $\bar{\alpha}=0.2$. 
$\bar{g}$ was choosen to be $0.6168576$.
}
\end{figure}
Directly related to these blocks is the behaviour of $\langle
t\rangle_i$, the mean number of topplings per site per unit time,
which we observed as a function of the site index $i$, averaged over a
long time and over many systems, as shown in figure \ref{figure7}.

The fact that sites in the synchronized regions topple the same number 
of times, while those at the boundaries topple less often
(due to the missing neighbours, or due to neighbours toppling less often),
can also be explained analytically in the following way: 
At every site, a local balance equation has to be fulfilled.
Let $\bar g$ be the mean force increment per unit time, 
and $\bar\alpha$ the mean package size, 
which we assume to be the same at each site.
The balance equation then is
\begin{equation}
\bar{g}+\bar{\alpha}\left(\langle t \rangle_{i-1}+\langle t \rangle_{i+1}\right)=\langle t \rangle_i \label{balance}
\end{equation}
for each site $i$, with the boundary condition 
${\langle t \rangle_0=\langle t \rangle_{L+1}=0}$. 
Eq. (\ref{balance}) can be written in matrix form
\begin{equation}
\langle\vec t\rangle_L=\bar{g}\mathbf{\Gamma}^{-1}_{L\times L}
\vec d_L\qquad,
\end{equation}
where $\mathbf{\Gamma}$ is tridiagonal and given by
\begin{equation}
\mathbf{\Gamma}_{L\times L}=
\left(
\begin{array}{ccccc}
1 &-\bar{\alpha} & 0 & \dots & 0 \cr
-\bar{\alpha} & 1 &-\bar{\alpha} & & \cr
\vdots&&\ddots &&\vdots \cr
&&-\bar{\alpha}&1&-\bar{\alpha}\cr
0&\dots&&-\bar{\alpha} &1
\end{array}\right)_{L\times L}
\end{equation}
and $\vec d_L$ is a vector with constant entries 1.  The numerical
solution of Eq. (\ref{balance}) is also shown in Figure
\ref{figure7}. An analytical solution can be found by making a
continuum approximation to Eq.~(\ref{balance}),
\begin{equation}
\frac{d^2 \langle t(x) \rangle}{dx^2} = \frac{1-2\bar \alpha}{\bar \alpha} 
\langle t(x) \rangle-\frac{\bar g}{\bar \alpha}\, .
\end{equation}
The solution that satisfies the boundary conditions is
\begin{eqnarray}
\langle t(x)\rangle&=&\frac{\bar g}{1-2\bar\alpha}\\
&&\hspace{-1cm}\left[
\frac{\sinh\left(\gamma(x-L-1)\right)}{\sinh\left(\gamma(L+1)\right)}-
\frac{\sinh\gamma x}{\sinh\left(\gamma(L+1)\right)}+1
\right]\nonumber
\end{eqnarray}
(with $\gamma=\frac{1-2\bar\alpha}{\bar\alpha}$).
This mean-field result predicts that the thickness of the boundary layer is
proportional to $\alpha$, which agrees with our previous numerical
finding that the time until the onset of synchronization is
proportional to $\alpha$.

\begin{figure}[htb]
\begin{center}
\mbox{}
\hspace*{-1cm}\includegraphics[height=6.2cm]{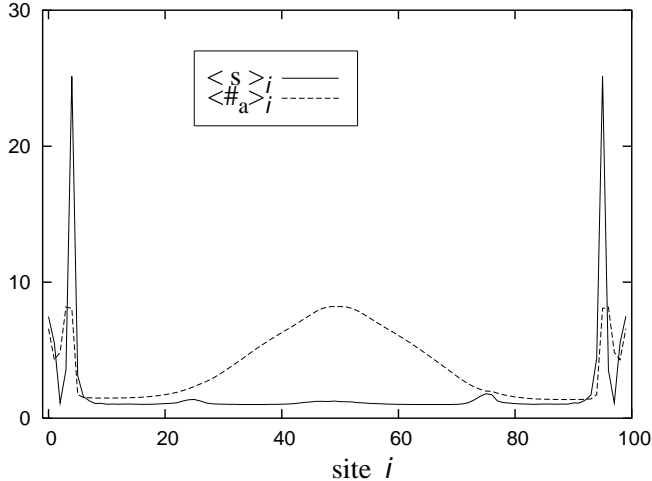}
\end{center}
\caption{\label{figure8} Solid line: $\langle s\rangle_i$: mean
size of avalanches triggered at site $i$; dashed line: $\langle
\#_a\rangle_i$: mean number of avalanches triggered at site $i$.  Both
data sets are plotted as function of the site index $i$, averaged over
10000 different synchronized systems for $\alpha=0.2$, $L=100$.  }
\end{figure}

We also considered $\langle s\rangle_i$, the mean size of all the
avalanches triggered at site $i$.  The result for $\alpha=0.2$ and
$L=100$, averaged over 10000 synchronized systems, is shown in figure
\ref{figure8}.  Almost all of the large avalanches are triggered
near the boundaries.  Also shown in figure \ref{figure8} is the
relative number of avalanches triggered at site $i$, which also shows
narrow peaks at the boundaries of the system, but also a broad peak in
the center.

Combining the two data sets, we arrive at the following scenario: In
the stationary state, most of the avalanches are single topplings. All
large avalanches are triggered near the boundaries and extend far into
the synchronized block. If they do not reach the end of the
synchronized block, the rest of the block topples in a series of
smaller avalanches, mostly of size 1. These small avalanches cause the
broad peak at the center of Figure \ref{figure8}.  The structure
of the peaks at the boundary of the curves in figure
\ref{figure8} depends on $\alpha$, and results from the
averaging over many different stationary states. 
\begin{figure}[htb]
\begin{center}
\mbox{}
\includegraphics[width=0.48\columnwidth]{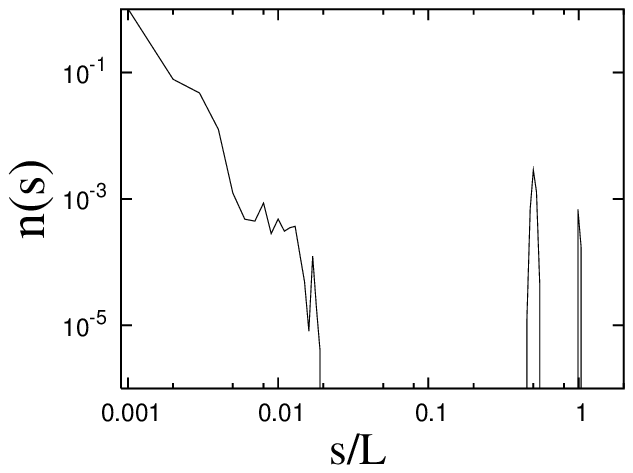}
\includegraphics[width=0.48\columnwidth]{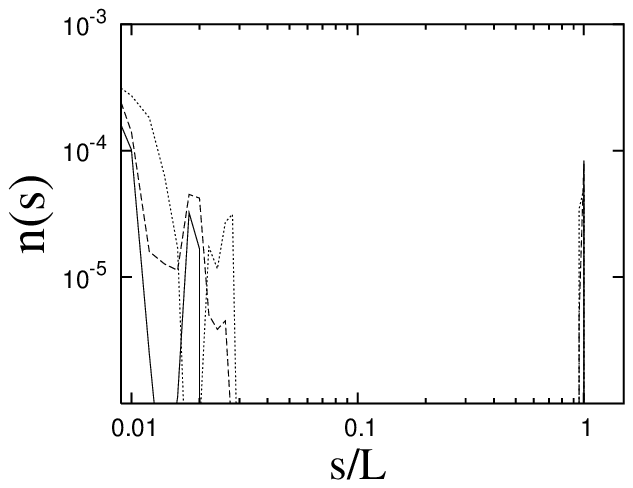}
\includegraphics[width=0.48\columnwidth]{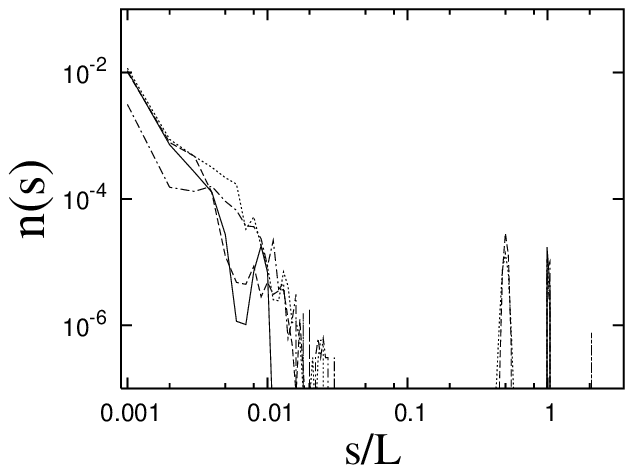}
\includegraphics[width=0.48\columnwidth]{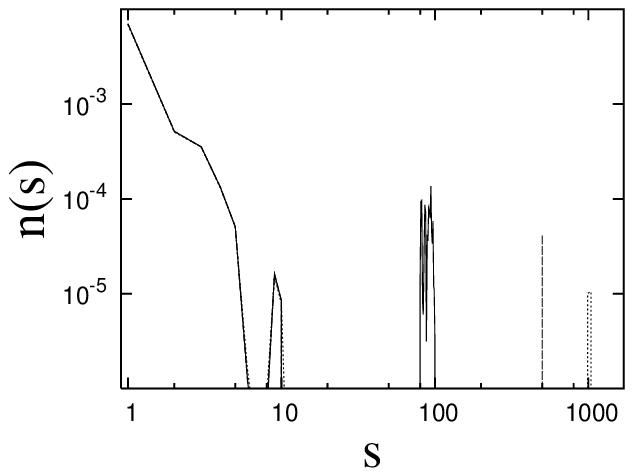}
\end{center}
\caption{\label{figure9} Number of avalanches $n(s)$ of size $s$, divided
by the total number of topplings and the system size, 
averaged over at least 2000 systems 
Top left: $L=1000, \alpha=0.2$.
Top right: $L=500, \alpha=0.15$;
solid line: precision $2^{-12}$; 
dashed line: precision $2^{-20}$; 
dotted line: precision $2^{-28}$. 
Bottom left: $L=1000$;
solid line: $\alpha=0.1$; 
dashed line: $\alpha=0.2$; 
dotted line: $\alpha=0.3$;
dash-dotted line: $\alpha=0.4$.
Bottom right: $\alpha=0.15$;
solid line: $L=100$; 
dashed line: $L=500$; 
dotted line: $L=1000$.}
\end{figure}

We now turn to the size distribution of the avalanches.  For
two-dimensional systems this is believed to obey a power law (see for
example \cite{lis01,lis01a}).  However, the one-dimensional systems
show this feature only for short times or for small system
sizes.  In the stationary state, the avalanche-size distribution 
looks like the upper left curve of figure \ref{figure9}.  It
was obtained by averaging over 2140 different systems and $10^9$
topplings, where we neglected the first $10^9$ transient topplings of
each system.

For small avalanche sizes, a power law is visible, but only for a
single decade and up to the sharp cutoff.  We see a large gap,
followed by peaks centered at system size and half the system size.
The shape of $n(s)$ for small $s$ depends only on the value of
$\alpha$ and on the precision used in the simulations, but not on $L$.
We checked this by inserting more sites into the synchronized blocks
and comparing the resulting avalanche-size distributions, which only
differed for the large avalanches. (See also the bottom right curve 
of figure \ref{figure9}). 

For larger precision, more avalanches are found. (See the upper right
curves of figure \ref{figure9}). The reason for this is that the
period of the stationary state is longer, as stated already before.
Smaller values of $\alpha$ result also in smaller avalanches, because
a higher precision is needed in order to resolve force differences
(which scale with powers of $\alpha$). Note also the nonvanishing
weight for avalanches of size $2L$ for $\alpha>\alpha_c$ in the lower
left figure.

The results for $n(s)$ confirm the picture that the system is composed
of a boundary layer that controls the dynamics and determines the
stationary state, and a synchronized block of sites that topple the
same number of times and that can be made larger without modifying the
boundaries.

\section{Discussion}
\label{conclusion}

The investigation of the one-dimensional version of the self-organized
critical earthquake models has revealed many intreaguing features. When
viewed as a dynamical system, the system shows 4 different types of
attractors, all of them being periodic. In contrast to a 2-site version
of the model, where the variables can only change continuously in time
\cite{sou99}, and to a many-site version, where the reset rule is $z_i
\to z_i-z_c$ \cite{cri92}, we do not find chaotic attractors.  In
contrast to the one-dimensional Zhang-model, which is conservative and
has a stochastic force input \cite{bla00}, the phase space volume does not
necessarily shrink for systems that have the same sequence of
topplings and avalanches.

In the stationary state, the model consists of two boundary layers, the
thickness of which is larger for longer attractors, and an inner part
consisting of one or two synchronized blocks, where all sites have
approximately the same force value. The synchronized blocks can be
made larger without changing the dynamics of the boundary layer or the
period of the attractor. Large avalanches are always triggered near
the boundary. These features are clearly reflected in the
avalanche-size distribution, where the small avalanches are
independent of the system size for sufficiently large systems,
while the large avalanches are proportional to it.

Several features of the one-dimensional model are very similar to the
two-dimensional model, while others differ. 
In both versions, large
avalanches are only triggered at the boundaries \cite{lis01a}, and
synchronization proceeds inwards according to a power-law in time
\cite{mid95,lise02}. The inner part is dominated by avalanches of size
1 \cite{gras94,bot97,dro02} even in the stationary state. The
computing precision affects the avalanche-size distribution
\cite{dro02}. However, while there are less large avalanches for
smaller computing precision in one dimension, there are more large
avalanches in two dimensions. That the inner part can be made larger
without changing the dynamics of the boundary region, must be a
special property of the one-dimensional system
due to the fact that the boundary of a synchronized block is merely a point
and that avalanches can propagate only along lines.
Nevertheless, 
the inner part is to some extent slaved to the boundary region also in two
dimensions. The exact interplay between the two still needs to be
clarified. However, we want to conjecture that in two dimensions the
avalanche-size distribution separates also in two parts, when the
system size is only made large enough. We expect the first part to
become essentially independent of the system size, while the second
part becomes proportional to it. However, in order to verify (or
falsify) this conjecture, larger and faster computer simulations are
needed than those that have been performed up to now. 

Thus, it appears that the OFC-earthquake model 
is not self-organized critical in the sense of exhibiting 
avalanches of all sizes.
Rather, the avalanche-size distribution results from the combined effect 
of several mechanisms, and only for sufficiently large system sizes 
do different types of avalanches become clearly separated in size.

This work was supported by the Deutsche Forschungsgemeinschaft (DFG) 
under the contract (Dr300/3-1).

\bibliography{/home/felix/geTeXtes/paper/first_own/ofc}

\begin{thebibliography}{15}
\expandafter\ifx\csname natexlab\endcsname\relax\def\natexlab#1{#1}\fi
\expandafter\ifx\csname bibnamefont\endcsname\relax
  \def\bibnamefont#1{#1}\fi
\expandafter\ifx\csname bibfnamefont\endcsname\relax
  \def\bibfnamefont#1{#1}\fi
\expandafter\ifx\csname citenamefont\endcsname\relax
  \def\citenamefont#1{#1}\fi
\expandafter\ifx\csname url\endcsname\relax
  \def\url#1{\texttt{#1}}\fi
\expandafter\ifx\csname urlprefix\endcsname\relax\def\urlprefix{URL }\fi
\providecommand{\bibinfo}[2]{#2}
\providecommand{\eprint}[2][]{\url{#2}}

\bibitem[{\citenamefont{Olami et~al.}(1992)\citenamefont{Olami, Feder, and
  Christensen}}]{ofc92}
\bibinfo{author}{\bibfnamefont{Z.}~\bibnamefont{Olami}},
  \bibinfo{author}{\bibfnamefont{H.~J.~S.} \bibnamefont{Feder}},
  \bibnamefont{and}
  \bibinfo{author}{\bibfnamefont{K.}~\bibnamefont{Christensen}},
  \bibinfo{journal}{Phys.\ Rev.\ Lett.} \textbf{\bibinfo{volume}{68}},
  \bibinfo{pages}{1244} (\bibinfo{year}{1992}).

\bibitem[{\citenamefont{Middleton and Tang}(1995)}]{mid95}
\bibinfo{author}{\bibfnamefont{A.~A.} \bibnamefont{Middleton}}
  \bibnamefont{and} \bibinfo{author}{\bibfnamefont{C.}~\bibnamefont{Tang}},
  \bibinfo{journal}{Phys.\ Rev.\ Lett.} \textbf{\bibinfo{volume}{74}},
  \bibinfo{pages}{742} (\bibinfo{year}{1995}).

\bibitem[{\citenamefont{Grassberger}(1994)}]{gras94}
\bibinfo{author}{\bibfnamefont{P.}~\bibnamefont{Grassberger}},
  \bibinfo{journal}{Phys.\ Rev.\ E} \textbf{\bibinfo{volume}{49}},
  \bibinfo{pages}{2436} (\bibinfo{year}{1994}).

\bibitem[{\citenamefont{Lise and Paczuski}(2001{\natexlab{a}})}]{lis01}
\bibinfo{author}{\bibfnamefont{S.}~\bibnamefont{Lise}} \bibnamefont{and}
  \bibinfo{author}{\bibfnamefont{M.}~\bibnamefont{Paczuski}},
  \bibinfo{journal}{Phys.\ Rev.\ E} \textbf{\bibinfo{volume}{63}},
  \bibinfo{pages}{036111} (\bibinfo{year}{2001}{\natexlab{a}}).

\bibitem[{\citenamefont{Lise and Paczuski}(2001{\natexlab{b}})}]{lis01a}
\bibinfo{author}{\bibfnamefont{S.}~\bibnamefont{Lise}} \bibnamefont{and}
  \bibinfo{author}{\bibfnamefont{M.}~\bibnamefont{Paczuski}},
  \bibinfo{journal}{Phys.\ Rev.\ E} \textbf{\bibinfo{volume}{64}},
  \bibinfo{pages}{046111} (\bibinfo{year}{2001}{\natexlab{b}}).

\bibitem[{\citenamefont{P\'erez et~al.}(1996)\citenamefont{P\'erez, Corral,
  D\'iaz-Guilera, Christensen, and Arenas}}]{per96}
\bibinfo{author}{\bibfnamefont{C.}~\bibnamefont{P\'erez}},
  \bibinfo{author}{\bibfnamefont{A.}~\bibnamefont{Corral}},
  \bibinfo{author}{\bibfnamefont{A.}~\bibnamefont{D\'iaz-Guilera}},
  \bibinfo{author}{\bibfnamefont{K.}~\bibnamefont{Christensen}},
  \bibnamefont{and} \bibinfo{author}{\bibfnamefont{A.}~\bibnamefont{Arenas}},
  \bibinfo{journal}{Int. J. Mod. Phys. B} \textbf{\bibinfo{volume}{10}},
  \bibinfo{pages}{1111} (\bibinfo{year}{1996}).

\bibitem[{\citenamefont{Mousseau}(1996)}]{mou96}
\bibinfo{author}{\bibfnamefont{N.}~\bibnamefont{Mousseau}},
  \bibinfo{journal}{Phys.\ Rev.\ Lett.} \textbf{\bibinfo{volume}{77}},
  \bibinfo{pages}{968} (\bibinfo{year}{1996}).

\bibitem[{\citenamefont{Drossel}(2002)}]{dro02}
\bibinfo{author}{\bibfnamefont{B.}~\bibnamefont{Drossel}},
  \bibinfo{journal}{Phys.\ Rev.\ Lett.} \textbf{\bibinfo{volume}{89}},
  \bibinfo{pages}{238701} (\bibinfo{year}{2002}).

\bibitem[{\citenamefont{Hergarten and Neugebauer}(2002)}]{her02}
\bibinfo{author}{\bibfnamefont{S.}~\bibnamefont{Hergarten}} \bibnamefont{and}
  \bibinfo{author}{\bibfnamefont{H.}~\bibnamefont{Neugebauer}},
  \bibinfo{journal}{Phys.\ Rev.\ Lett.} \textbf{\bibinfo{volume}{88}},
  \bibinfo{pages}{238501} (\bibinfo{year}{2002}).

\bibitem[{\citenamefont{Burridge and Knopoff}(1967)}]{bur67}
\bibinfo{author}{\bibfnamefont{R.}~\bibnamefont{Burridge}} \bibnamefont{and}
  \bibinfo{author}{\bibfnamefont{L.}~\bibnamefont{Knopoff}},
  \bibinfo{journal}{Bull.\ Seismol.\ Soc.\ Am.} \textbf{\bibinfo{volume}{57}},
  \bibinfo{pages}{341} (\bibinfo{year}{1967}).

\bibitem[{\citenamefont{de~Sousa~Vieira}(1999)}]{sou99}
\bibinfo{author}{\bibfnamefont{M.}~\bibnamefont{de~Sousa~Vieira}},
  \bibinfo{journal}{Phys.\ Rev.\ Lett.} \textbf{\bibinfo{volume}{82}},
  \bibinfo{pages}{201} (\bibinfo{year}{1999}).

\bibitem[{\citenamefont{A.~Crisanti and Paladin}(1992)}]{cri92}
\bibinfo{author}{\bibfnamefont{A.~V.} \bibnamefont{A.~Crisanti},
  \bibfnamefont{M.H.~Jensen}} \bibnamefont{and}
  \bibinfo{author}{\bibfnamefont{G.}~\bibnamefont{Paladin}},
  \bibinfo{journal}{Phys.\ Rev.\ A} \textbf{\bibinfo{volume}{46}},
  \bibinfo{pages}{R7363} (\bibinfo{year}{1992}).

\bibitem[{\citenamefont{P.~Blanchard and Kruger}(2000)}]{bla00}
\bibinfo{author}{\bibfnamefont{B.~C.} \bibnamefont{P.~Blanchard}}
  \bibnamefont{and} \bibinfo{author}{\bibfnamefont{T.}~\bibnamefont{Kruger}},
  \bibinfo{journal}{J.\ Stat.\ Phys.} \textbf{\bibinfo{volume}{98}},
  \bibinfo{pages}{375} (\bibinfo{year}{2000}).

\bibitem[{\citenamefont{Lise}(2002)}]{lise02}
\bibinfo{author}{\bibfnamefont{S.}~\bibnamefont{Lise}},
  \bibinfo{journal}{cond-mat/0204490}  (\bibinfo{year}{2002}).

\bibitem[{\citenamefont{Bottani and Delamotte}(1997)}]{bot97}
\bibinfo{author}{\bibfnamefont{S.}~\bibnamefont{Bottani}} \bibnamefont{and}
  \bibinfo{author}{\bibfnamefont{B.}~\bibnamefont{Delamotte}},
  \bibinfo{journal}{Physica D} \textbf{\bibinfo{volume}{103}},
  \bibinfo{pages}{430} (\bibinfo{year}{1997}).

\end{thebibliography}
\end{document}